\PassOptionsToPackage{table}{xcolor}
\documentclass[conference]{IEEEtran}
\IEEEoverridecommandlockouts
\usepackage{cite}
\usepackage{amsmath,amssymb,amsfonts}
\usepackage{algorithmic}
\usepackage{graphicx}
\usepackage{textcomp}
\usepackage{xcolor}

\usepackage[nolist]{acronym}
\usepackage[table]{xcolor}  % Load xcolor with table option
\usepackage{multirow}
\usepackage{tabularx}
\usepackage{hhline}

\usepackage{booktabs}    % For professional-looking tables
\usepackage{colortbl}    % To color table rows properly
\usepackage{array}       % For better column formatting
\usepackage{enumerate}
\usepackage{xspace}
\usepackage{tikz}

\newcommand{\mypara}[1]{\vspace{2pt}\noindent{\textit{\textbf{#1}}}\xspace}

\newcommand*\circled[1]{\tikz[baseline=(char.base)]{
            \node[shape=circle,draw,inner sep=2pt] (char) {#1};}}

\begin{acronym}
    \acro{COTS}{Commercial Off-The-Shelf}
    \acro{GPUs}{Graphics Processing Units}
    \acro{TPUs}{Tensor Processing Units}
    \acro{NPUs}{Neural Processing Units}
    \acro{CPUs}{Central Processing Units}
    \acro{FPGAs}{Field-Programmable Gate Arrays}
    \acro{SWaP-C}{Size, Weight, Power, and Cost}
    \acro{MCSs}{Mixed Criticality Systems}
    \acro{SPH}{Static Partitioning Hypervisors}
    \acro{LLC}{Last-Level Cache}
    \acro{WCET}{Worst-Case Execution Time}
    \acro{VMs}{Virtual Machines}
    \acro{MBR}{Memory Bandwidth Reser}
    \acro{SMMU}{System Memory Management Unit}
    \acro{IOMMU}{Input/Output Memory Management Unit}
    \acro{IOTLB}{Input/Output Translation Lookaside Buffer}
    \acro{TBU}{Translation Buffer Unit}
    \acro{TLB}{Translation Lookaside Buffer}
    \acro{DMA}{Direct Memory Access}
    \acro{FPD}{Full-Power Domain}
    \acro{LPD}{Low-Power Domain}
    \acro{PMU}{Performance Monitor Unit}
    \acro{ILA}{Integrated Logic Analyzer}
    \acro{FPGA}{Field-Programmable Array}
    \acro{PTW}{Page Table Walk}
    \acro{TCU}{Translation Control Unit}
    \acro{SoC}{System on Chip}
    \acro{MCUs}{Microcontroller Units}
\end{acronym}

\def\BibTeX{{\rm B\kern-.05em{\sc i\kern-.025em b}\kern-.08em
    T\kern-.1667em\lower.7ex\hbox{E}\kern-.125emX}}
\begin{document}

\title{Beyond the Bermuda Triangle of Contention: IOMMU Interference in Mixed Criticality Systems\\

%\thanks{\textcolor{red}{Identify applicable funding agency here. If none, delete this.}}
}

\author{\IEEEauthorblockN{1\textsuperscript{st} Diogo Costa}
\IEEEauthorblockA{\textit{Centro ALGORITMI / LASI} \\
\textit{Universidade do Minho}\\
%City, Country \\
id10560@alunos.uminho.pt}
\and
\IEEEauthorblockN{2\textsuperscript{nd} José Martins}
\IEEEauthorblockA{\textit{Centro ALGORITMI / LASI} \\
\textit{Universidade do Minho}\\
%City, Country \\
jose.martins@dei.uminho.pt}
\and
\IEEEauthorblockN{3\textsuperscript{rd} Sandro Pinto}
\IEEEauthorblockA{\textit{Centro ALGORITMI / LASI} \\
\textit{Universidade do Minho}\\
%City, Country \\
sandro.pinto@dei.uminho.pt}
}

\maketitle

\begin{abstract} 
As \ac{MCSs} evolve, they increasingly integrate heterogeneous computing platforms, combining general-purpose processors with specialized accelerators such as AI engines, GPUs, and high-speed networking interfaces. This heterogeneity introduces challenges, as these accelerators and DMA-capable devices act as independent bus masters, directly accessing memory. Consequently, ensuring both security and timing predictability in such environments becomes critical.
To address these concerns, the Input-Output Memory Management Unit (IOMMU) plays a key role in mediating and regulating memory access, preventing unauthorized transactions while enforcing isolation and access control policies. While prior work has explored IOMMU-related side-channel vulnerabilities from a security standpoint, its role in performance interference remains largely unexplored. Moreover, many of the same architectural properties that enable side-channel leakage, such as shared TLBs, caching effects, and translation overheads, can also introduce timing unpredictability. 
In this work, we analyze the contention effects within IOMMU structures using the Xilinx UltraScale+ ZCU104 platform, demonstrating how their shared nature introduce unpredictable delays. Our findings reveal that IOMMU-induced interference primarily affects small memory transactions, where translation overheads significantly impact execution time. Additionally, we hypothesize that contention effects arising from IOTLBs exhibit similar behavior across architectures due to shared caching principles, such as prefetching and hierarchical TLB structures. Notably, our experiments show that IOMMU interference can delay DMA transactions by up to 1.79× for lower-size transfers on the Arm SMMUv2 implementation. 
%This highlights the need to account for DMA transactions and the impact of IOMMU translation overhead on their expected timing behavior when designing time-sensitive MCS workloads.}

\end{abstract}


\section{Introduction}

In recent decades, the trend toward digitization \cite{cerrolaza2020multi, LTZVisor2017} has reshaped numerous industries, including automotive, robotics, and aerospace. This evolution has rapidly increased in system complexity, with high-end embedded platforms evolving from simple, single-core \ac{MCUs} into complex, heterogeneous architectures \cite{costa2023}. These modern systems integrate multi-core CPUs and specialized hardware accelerators such as \ac{GPUs}, \ac{TPUs}, \ac{NPUs}, and \acp{FPGA} \cite{gracioli2019, mancuso2013real}, enabling unprecedented levels of performance and functionality.
At the same time, the demand for integration and efficiency has driven the consolidation of multiple workloads onto a single hardware platform, as integrating peripherals and computational tasks helps meet stringent \ac{SWaP-C} requirements. This shift gave rise to Mixed Criticality Systems (MCSs), where systems with different criticality levels co-exist on the same hardware \cite{henzinger2006embedded}.

Consolidating mixed-criticality workloads onto a shared platform presents significant challenges in ensuring safety, security, and isolation \cite{cinque2022virtualizing, costa2023}. To address these challenges, virtualization has emerged as a key technology. \ac{SPH} \cite{bao2020, klein2009sel4, hwang2008xen, jailhouse2017} represent the zeitgeist of \ac{MCSs} hypervisors, as they leverage static resource allocation to ensure isolation and determinism \cite{martins2023shedding, ottaviano2024}.
While \ac{SPH}s effectively enforce spatial isolation by partitioning architectural resources (e.g. memory regions and devices) among \ac{VMs}, they cannot fully guarantee temporal isolation, as microarchitectural resources, including the \ac{LLC}, main memory, and system bus, remain inherently shared. The resulting resource contention introduces two key challenges: (i) increased execution time and (ii) reduced determinism, caused by unpredictable delays \cite{oliveira_2024, yun2024, costa2022, cazorla2019probabilistic, Tamara2022}. These timing variations pose a significant challenge in hard real-time systems, where strict \ac{WCET} guarantees are essential to ensuring system correctness and safety \cite{abella2015wcet, cazorla2019probabilistic}.

Over the past decade, considerable research has been conducted to identify and mitigate interference caused by shared micro-architectural resources. The majority of this work has focused on three primary sources of contention: (i) the \ac{LLC} \cite{Gracioli2015, oliveira2024ia, Bechtel2019}, (ii) the system bus \cite{zuepke2024mempol, kotaba2013multicore, dasari2013identifying, kloda2019deterministic}, and (iii) the main memory \cite{zuepke2024mempol, Izhbirdeev2024, lofwenmark2016understanding, MemGuard, Yun2014, yun2012memory, yun2012memory} - forming what we refer to as the \textit{'Bermuda Triangle of Contention'}. While these sources have been well studied, modern platforms incorporate a broader array of shared resources (e.g., interrupt controllers \cite{costa2023}), whose effects on contention remain insufficiently explored.

\begin{figure*}[t]
    \centering
    \includegraphics[width=\linewidth]{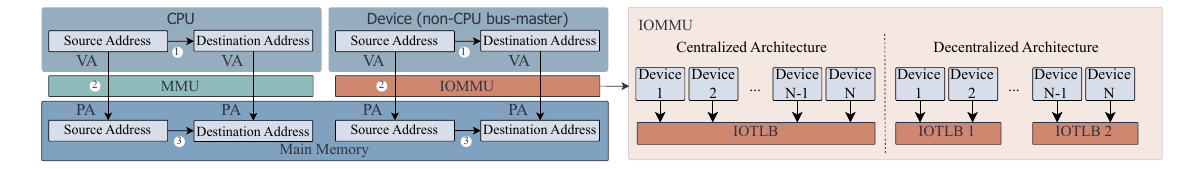}
   \caption{Conceptual overview of address translation for CPU and non-CPU bus-master devices, comparing MMU and IOMMU operations.}
    \label{fig:IOMMU_functionality_overview}
\end{figure*}

Among these overlooked components, the \ac{IOMMU} plays a critical role in enforcing memory isolation for non-CPU bus masters, preventing unauthorized memory access and enhancing system security \cite{morgan2018iommu, tang2016iommu}. While IOMMUs are primarily recognized for these security functions, existing research has focused on a narrow set of concerns, mainly their susceptibility to side-channel attacks such as \ac{IOTLB}-based timing attacks \cite{morgan2018iommu, zhu2017understanding}. In contrast, the implications of IOMMU design and usage on timing predictability in consolidated \ac{MCSs} remain largely unexplored. In such systems, where multiple devices may simultaneously share IOMMU resources, contention in translation structures such as address translation overheads and IOTLB cache pressure can lead to unpredictable latency, undermining the determinism required for real-time applications.

This oversight is concerning, as such contention not only impacts system behavior but can also undermine timing determinism in real-time workloads. Despite the critical role of the \ac{IOMMU} in \ac{MCSs}, there has been little investigation into how contention for its internal resources may affect execution latency and predictability.
In this paper, we aim to advance the understanding of these effects through an empirical study of contention in a representative \ac{COTS} platform: the Xilinx UltraScale+ ZCU104. This platform integrates an \ac{IOMMU} compliant with the ARM SMMU-500 (SMMUv2) specification. a decentralized architecture in which multiple Translation Buffer Units (TBUs) handle address translation for device clusters. We adapt established microbenchmarking techniques to the IOMMU domain to quantify the impact of shared translation structures under concurrent DMA traffic. Our contributions are as follows:

\begin{enumerate}
    \item \textbf{Reverse-Engineering IOMMU Structures for Contention Analysis} – In this work, we reverse-engineer the translation structures of the IOMMU implementation found in the ZCU104 platform. By uncovering the microarchitectural details of the IOMMU, we gain valuable insights into how translation caches, like the IOTLB, contribute to contention and DMA performance degradation. This reverse-engineering effort enhances our understanding of IOMMU internals and provides procedures that can be applied in future research to analyze IOMMU contention across other platforms and architectures.

    \item \textbf{Characterizing Contention Effects on Translation Latency} – Our experimental results show that IOMMU contention can introduce measurable latency increases (up to 1.79×) in DMA transactions. While moderate in magnitude, these delays may affect fine-grained timing behavior in real-time systems. We discuss these effects from a microarchitectural perspective and defer broader system-level implications to future work.
    
    \item \textbf{Open-Source Artifacts for Independent Validation and Further Exploration} – To foster further research and validation, we make all experimental artifacts, including code, configurations, and data, publicly available to the research community. enabling independent validation and serving as a foundation for further research on IOMMU behavior across other hardware and system architectures.

\end{enumerate}

\section{Background}

Efficient \ac{DMA} has become essential in modern computing, allowing high-speed data transfers and improving overall system performance. Devices with DMA capabilities, such as hardware accelerators, rely on direct access to system memory, improving data processing speed and operational efficiency \cite{rossi2014ultra}. However, this direct access raises concerns about security and data integrity, particularly in consolidated computing environments where multiple systems share the same hardware. Ensuring both workload-specific isolation and system-wide protection requires robust mechanisms to control and restrict memory access. IOMMUs are pivotal to protect against such threats.

As shown on the left side of Figure~\ref{fig:IOMMU_functionality_overview}, when a CPU issues a memory request, it uses a Virtual Address (VA) \circled{1}, which the MMU translates to a Physical Address (PA) \circled{2} by consulting its page tables, typically via a \ac{PTW}. This PA is then used to access the appropriate location in main memory \circled{3}. Similarly, on the right side, when a non-CPU bus master (e.g., a DMA-capable device) initiates a memory transaction, it uses a virtual address \circled{1} within its assigned I/O address space. This address is intercepted by the IOMMU, which performs the translation to a physical address \circled{2}, accessing the I/O page tables via a \ac{PTW} to enforce isolation and access control policies. The resulting PA is used to access the corresponding memory location \circled{3}. To perform these translations efficiently, the IOMMU maintains I/O page tables, analogous to those used by the MMU. Additionally, it uses \ac{IOTLB}s to cache recently translated address mappings and reduce translation latency. The right side of Figure~\ref{fig:IOMMU_functionality_overview} highlights two architectural models: (i) a centralized design, where a single shared IOTLB serves all devices; and (ii) a decentralized approach, where multiple IOTLBs are distributed across device clusters. These designs offer trade-offs in latency, scalability, and isolation granularity. By enforcing strict access control and isolating memory domains across devices, the IOMMU significantly enhances system reliability and security.

While the security benefits of \ac{IOMMU}s are well-documented, their impact on performance remains an underexplored area, particularly in \ac{MCSs}. The overhead introduced by address translation, permission enforcement, and IOTLB management can create latency and contention when multiple devices share the same IOMMU. This contention is especially concerning for real-time and safety-critical applications, where predictable timing is crucial. To provide context for our study, we briefly review the ARM \ac{SMMU} architecture, which serves as the reference IOMMU implementation in our evaluation. While other specifications exist, including Intel VT-d \cite{intel-vtd}, AMD-Vi \cite{amd-vi}, and the RISC-V IOMMU \cite{riscv-iommu}, this work focuses specifically on ARM’s specification. The ARM SMMUv2 \cite{arm-SMMUv2} is the standard IOMMU implementation for arm64 systems, providing secure memory isolation and efficient address translation. It employs a stream-based model, using StreamIDs and optional SubstreamIDs for fine-grained device identification. SMMUv2 supports both stage-1 and stage-2 address translations and integrates tightly with ARM’s memory architecture. To improve lookup efficiency, it integrates IOTLBs at the interface between devices and the SMMU, reducing memory access overhead. Additionally, it defines three circular buffer queues for translation updates, fault handling, and page requests.

IOTLBs play a crucial role in accelerating address translation and reducing memory overhead, but their organization varies across IOMMU implementations. Centralized designs, such as those in Intel VT-d and AMD-Vi, simplify caching but can become bottlenecks under high traffic. In contrast, architectures like Arm’s CoreLink MMU-500 and MMU-600, as well as the RISC-V IOMMU open-source implementation from Zero-Day Labs \cite{riscv-iommu-zeroday}, support more flexible or fully distributed IOTLB configurations. While decentralized designs improve scalability and reduce lookup latency, they also introduce challenges in maintaining coherence and avoiding contention. These issues are not limited to specialized systems but extend to a wide range of \ac{COTS} platforms. With growing adoption of distributed IOMMUs (e.g., Arm forecasts over 330 million MMU-600-based systems \cite{arm-mmu600-savings}), contention due to frequent invalidations, synchronization overhead, and I/O-intensive workloads becomes a critical concern. This is particularly relevant in \ac{MCSs}, where timing predictability is essential, and in virtualized environments where devices assigned to different VMs compete for translation resources. Although this work focuses on a specific platform, our findings offer broader insight into the performance and scalability trade-offs posed by modern IOMMU designs.

\section{Methodology and Experimental Setup}

This section presents our methodology for evaluating IOMMU contention. Our goal is to determine whether IOTLB resources are a significant source of contention and how their capacity affects translation overhead under varying workloads. To do this, we use micro-benchmarking and performance monitoring to analyze cache behavior and identify performance limitations in modern IOMMU designs.

\begin{figure*}[t]
    \centering
    \includegraphics[width=.92\linewidth]{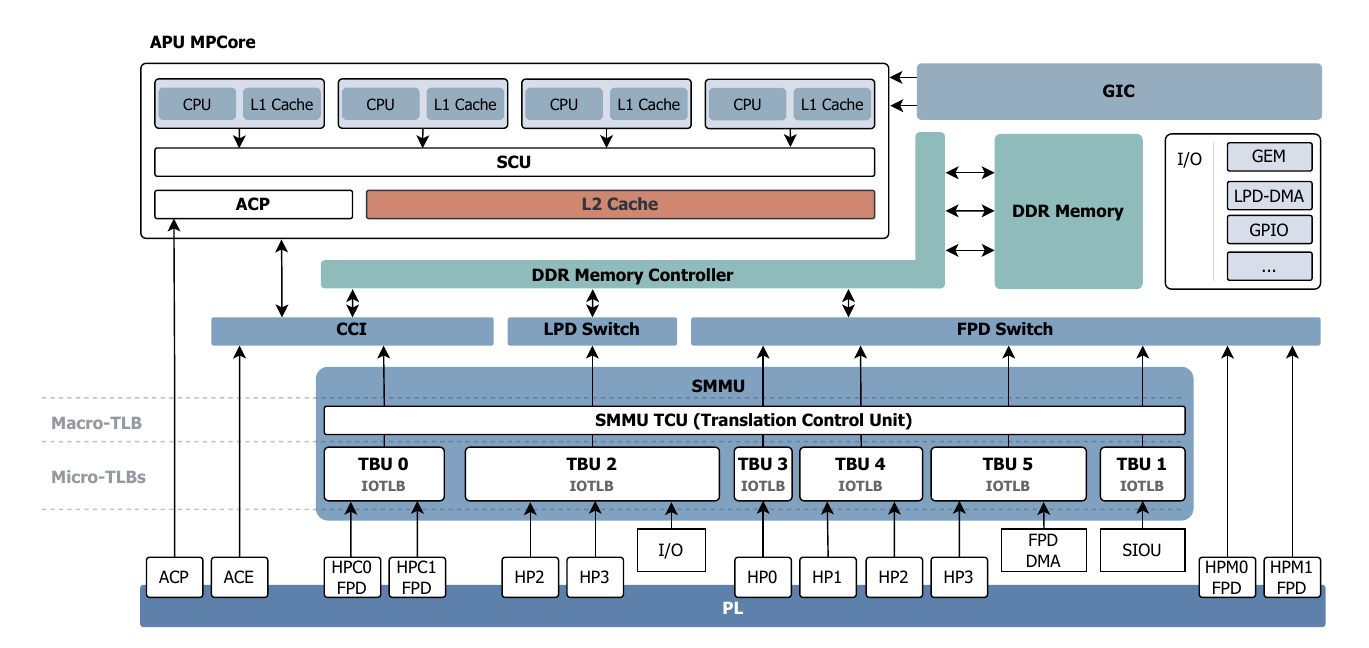}
    \caption{ZCU104 Platform Overview. The IOMMU implementation on this platform follows a decentralized architecture in which multiple \ac{TBU}s, functionally equivalent to IOTLBs, are associated with clusters of devices. These TBUs, also referred to as micro-TLBs, are part of the SMMU’s Translation Control Unit (TCU) caching subsystem, which also includes a centralized macro-TLB.}
    \label{fig:platform_overview}
\end{figure*}

\mypara{Hardware Platform.} Experiments were carried out on the Xilinx ZCU104 evaluation board, which integrates a Zynq UltraScale+ \ac{SoC} and serves as a representative platform for studying IOMMU architectures. It features a quad-core Arm Cortex-A53 processor (1.2 GHz) and the Arm CoreLink MMU-500, an implementation of SMMUv2 running at 525 MHz. The SoC also includes an \ac{FPGA} (100 MHz) and two \ac{DMA} engines: one in the \ac{FPD} (600 MHz) and one in the \ac{LPD} (500 MHz). While our analysis targets this specific platform, the methodology is broadly applicable to systems that include: (i) a programmable device capable of issuing DMA requests, (ii) an IOMMU with accessible performance monitoring events, and (iii) a system configuration where the programmable interface and the interfering DMA share the same IOTLB.

\mypara{Measurement Tools.} To analyze the architectural behavior of the SMMU implemented in the Xilinx UltraScale+ ZCU104 platform, we leveraged the Arm SMMUv2 \ac{PMU} to collect micro-architectural events, which enables the measurement of (i) cycle counts, (ii) the number of \ac{TLB} allocations, and (iii) the number of \ac{SMMU} transactions processed. 
These measurements are used exclusively for characterizing the IOTLB structures and do not impact the execution timing of benchmarks.
To evaluate the performance impact of \ac{SMMU} contention, we deployed a DMA unit within the FPGA and leveraged the \ac{ILA} to obtain cycle-accurate execution times of \ac{DMA} transactions, capturing detailed hardware behavior without adding runtime overhead.

\mypara{Arm SMMUv2 Implementation on the ZCU104.}
The micro-architectural details of the SMMU on the ZCU104 platform are not publicly documented. To address this gap, we conducted reverse-engineering experiments to uncover key aspects of its implementation. According to the Arm SMMUv2 specification, the architecture features a hierarchical caching system designed to optimize address translation and memory management. The primary cache structures include:

\begin{enumerate}
\item \textbf{Micro-TLB}: Caches \ac{PTW} results returned by the \ac{TCU}.
\item \textbf{Macro-TLB and PTW Cache}: Caches \ac{PTW} results within the \ac{TCU}, with the \ac{PTW} cache specifically storing partial \acp{PTW} to reduce the frequency of full \acp{PTW}.
\item \textbf{Prefetch Buffer (IPA to PA Cache)}: Prefetches pages to minimize future \acp{PTW}.
\end{enumerate}
While the specification outlines general guidelines, such as a fully associative micro-\ac{TLB} and four-way set associative macro-\ac{TLB}, while the exact depth (number of entries) of these caches is vendor-specific. This configurability introduces challenges in evaluating the impact of shared structures within the \ac{SMMU}. To address these challenges, we developed a set of micro-benchmarks to empirically determine the cache depths of the MMU-500 implemented on this platform. According to the MMU-500 specification, the micro-\ac{TLB} can support up to 128 entries, backed by a \ac{TCU} cache with up to 2000 entries.

\mypara{Cache Depth Measurements.} To understand the impact of IOTLB contention, it is essential to investigate the cache depths within the Arm SMMU. The number of cache entries directly influences how quickly these caches will be evicted during high-frequency DMA workloads. Specifically, a smaller number of cache entries results in more frequent cache misses, which increases the overhead of address translation. Moreover, determining the precise cache depth allows us to quantify the number of memory pages that can be accessed while the translations from VAs to PAs remain cached. This information is crucial for understanding the scalability limitations of the IOTLB and identifying the point at which performance degradation begins to occur. While techniques for analyzing cache and TLB behavior via microbenchmarking are well-established in the real-time research community \cite{wilhelm2008worst}, particularly for uncovering \ac{WCET}, our work extends these methods to the IOMMU context. Specifically, we focus on the previously underexplored IOMMU components, such as the micro-TLB, in the context of modern heterogeneous platforms.
%\textcolor{red}{Regarding the timing behavior, our measurements do not indicate any meaningful difference between read and write operations in terms of how they interact with the IOMMU structures. This is because each DMA transaction inherently includes both a read and a write phase, and the IOTLB allocates translation entries in the same way for both. Therefore, our methodology focuses on determining the number of translation entries available (i.e., the cache depth) rather than timing differences between reads and writes, which were not observed to differ in our experimental setup.}
\label{sec:cache_depth_measurements}
To explore the behavior of the IOTLB and infer the available translation resources, we conducted a series of experiments to probe the cache depths of the SMMU on the ZCU104 platform. Specifically, we targeted the micro-\ac{TLB} and the macro-\ac{TLB}, that includes the PTW cache, by leveraging the SMMU PMU to monitor the number of translations processed and the associated allocations to the micro- and macro-TLBs. However, there is a limitation in the current setup: the SMMU's PMU only allows us to measure micro-architectural events related to the micro-TLBs, preventing a complete analysis of the macro-TLB configuration. As a result, we could only measure the depth of the micro-TLB, while for the macro-TLB, we relied on the depth of 2000 entries defined in the SMMUv2 specification.
To probe the micro-TLB depth, we leveraged the \ac{PMU} available in each TBU of the \ac{SMMU}. Specifically, we configured TBU5’s \ac{PMU} to monitor the following events: 
\begin{enumerate} 
    \item \textbf{Number of DMA transactions} processed by \ac{TBU}5 (read+write operations) — $SMMU\_access\ [T]$. 
    \item \textbf{Number of TLB entry allocations} triggered for DMA reads — $TLB\_entry\_alloc\ [R]$. 
    \item \textbf{Number of \ac{TLB} entry allocations} triggered for DMA writes — $TLB\_entry\_alloc\ [W]$. 
\end{enumerate}

\begin{figure*}[t]
\centering
\includegraphics[width=1\linewidth]{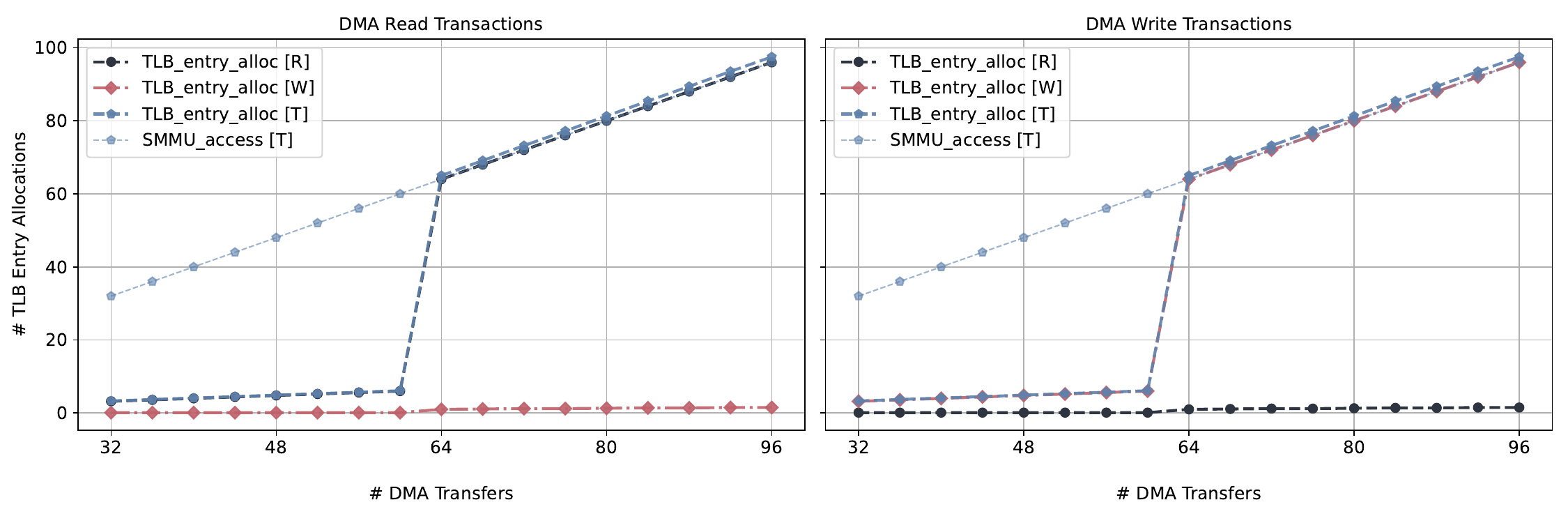}
\caption{SMMU TBU Micro-Benchmarking }
\label{fig:smmu_instr_evaluation}
\end{figure*}

To induce \ac{SMMU} translations and probe the behavior of the \ac{TBU}, we deployed a baremetal VM atop the Bao hypervisor. The baremetal's workload consists of a micro-benchmarking designed to leverage \ac{DMA} channels to perform memory transactions across different virtual memory pages. The micro-benchmarking runs (i) read and (ii) write operations: reads fetch small data chunks (e.g., 16B) from $N$ memory pages into a buffer, while writes store the buffer’s data into $N$ memory pages. These micro-benchmark leverages the \ac{FPD}-\ac{DMA} unit connected to \ac{TBU}5, as depicted in Figure~\ref{fig:platform_overview}.

Empirical results, presented in Figure \ref{fig:smmu_instr_evaluation}, reveal a linear correlation between the number of \ac{DMA} transactions and the corresponding \ac{SMMU} and \ac{TLB} accesses. For both read and write transactions, $SMMU\_access\ [T]$ increases proportionally with the number of \ac{DMA} transfers, demonstrating consistent handling of translation requests as the workload scales.
A key observation emerges when analyzing \ac{TLB} entry allocations, $TLB\_entry\_alloc\ [T]$. Before collecting micro-architectural events, the micro-benchmark runs multiple times to warm up the caches. This ensures that the required translations are already cached when measurements begin, resulting in zero \ac{TLB} allocations for smaller values of $N$.
However, as the number of DMA transactions increases, so does the number of \ac{TLB} allocations. This increase occurs when the number of transactions exceeds the capacity of the micro-\ac{TLB}.
The results from both read and write micro-benchmarking confirm that the micro-\ac{TLB} has a depth of 64 entries. However, the observed increase in $TLB\_entry\_alloc\ [T]$ occurs slightly earlier than expected—around the 64th transaction. This behavior is explained by the additional translation required to complete the DMA transfer. For example, in the read benchmark, an extra translation is needed for the write operation that stores the fetched data into the destination buffer. When performing 64 transactions, a total of 65 translations are needed, exceeding the micro-\ac{TLB} capacity and triggering allocations to the macro-\ac{TLB} for subsequent transactions. In contrast, with 63 transactions, the number of required translations precisely matches the cache size, resulting in no misses and no additional \ac{TLB} entry allocations. Our measurements show no significant difference in timing behavior between read and write operations in their interaction with the IOMMU structures. This is because each DMA transaction consists of both a read and a write phase, and the IOTLB allocates translation entries in the same manner for both. As a result, our methodology primarily focuses on determining the number of available translation entries, or the cache depth.

\section{IOMMU Contention and Performance Impact}

The \ac{IOMMU} plays a critical role in modern \ac{SoC} platforms by ensuring secure and controlled access to memory for \ac{DMA}-capable devices, thereby enhancing system security and reliability. However, the caching mechanisms in the \ac{IOMMU} introduce challenges related to both performance and predictability, particularly when multiple devices compete for shared resources.
In the case of the Arm \ac{SMMU}v2 used in our target platform, we identify two primary sources of contention: (i) the \ac{TBU}s (IOTLBs) distributed across the platform and (ii) the cache hierarchy within the \ac{SMMU}'s \ac{TCU}.

\begin{figure}[t]
    \centering
    \includegraphics[width=1\linewidth]{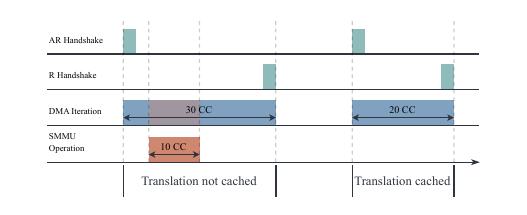}
    \caption{Toy example illustrating two DMA transactions in clock cycles (CC). The first corresponds to scenario (iii), where the translation is not cached, leading to higher latency. The second represents scenario (i), where the translation is cached, significantly reducing translation overhead.}
    \label{fig:smmu_contention_theo}
\end{figure}

\begin{figure}[t]
    \centering
    \includegraphics[width=1\linewidth]{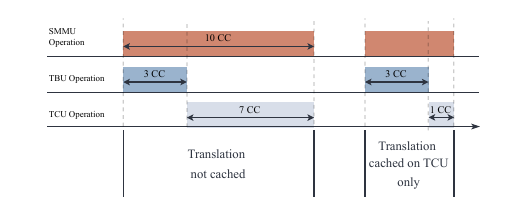}
    \caption{Toy example showing the impact of TBU and TCU caching effects on two DMA transactions. The first represents scenario (iii), where the translation is not cached in either the micro- or macro-TLB, leading to the highest latency. The second corresponds to scenario (ii), where the translation is absent from the micro-TLB but cached in the macro-TLB, reducing translation overhead.}
    \label{fig:smmu_contention_theo_diff_cache}
\end{figure}

\begin{figure*}[t]
    \centering
    \includegraphics[width=\linewidth]{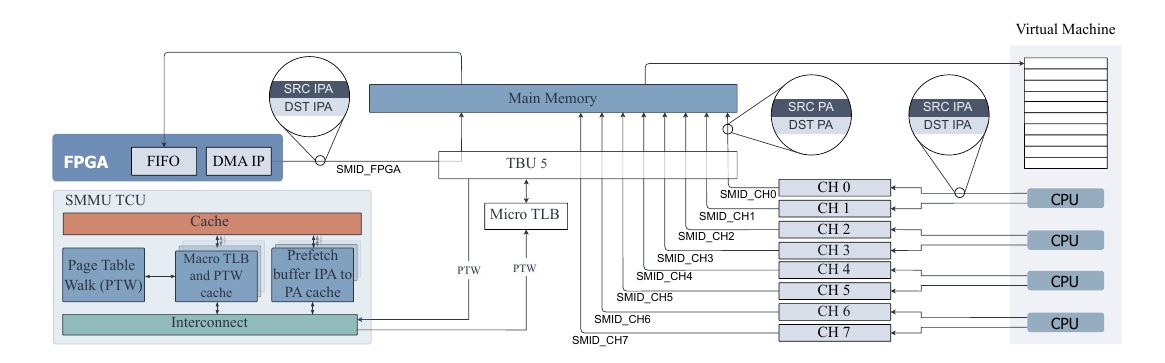}
    \caption{IOMMU Interference: Evaluation Setup}
    \label{fig:iommu_contention_setup}
\end{figure*}

As depicted in Figure~\ref{fig:smmu_contention_theo}, when a \ac{DMA} device initiates a memory transfer (\textit{\ac{DMA} iteration}), the process begins with an \textit{Address Read (AR)} handshake. The \ac{SMMU} then checks the address translation. If the translation is \textbf{not cached}, the \ac{SMMU} incurs a delay to perform a \ac{PTW} (if necessary) and fetch the translation into the \ac{TBU}. Conversely, if the translation is already \textbf{ cached}, the \ac{TBU} simply verifies the StreamID of the \ac{DMA} device and the VAs, and resumes the transaction with the correct PAs, with negligible overhead. This distinction is highlighted in Figure~\ref{fig:smmu_contention_theo}, where the latency differences between cached and non-cached translations are depicted. Further insights into the operation of the \ac{SMMU} are provided in Figure~\ref{fig:smmu_contention_theo_diff_cache}, which illustrates the detailed interaction between the \ac{TBU} and \ac{TCU}. When a translation is \textbf{not cached}, the process can be divided into two main stages: (i) the TBU stage and (ii) the TCU stage. In the TBU Stage: The \ac{TBU} first checks the \textit{micro-TLB} to determine if the required translation is cached locally. If it is not, the \ac{TBU} sends a request to the \ac{TCU} for further action. In the TCU Stage, upon receiving the request, the \ac{TCU} verifies if the translation is cached in the \textit{macro-TLB}. If it does, the translation is loaded into the \textit{micro-TLB} to complete the process. If not, the \ac{SMMU} performs a \ac{PTW}, fetches the required data, and updates the caches accordingly. As depicted in Figure~\ref{fig:smmu_contention_theo_diff_cache}, three distinct scenarios arise depending on the caching status of the translation:

\begin{enumerate}[(i)]
    \item \textbf{Micro-TLB hit} – The translation is cached directly in the \textit{micro-TLB}, allowing for near-instant translation and minimal latency.
    
    \item \textbf{Micro-TLB miss and Macro-TLB hit} – The translation is cached in the \textit{macro-TLB} but not in the \textit{micro-TLB}. Here, the overhead is lower, as the translation only needs to be fetched into the \textit{micro-TLB}.
    
    \item \textbf{Micro-TLB and Macro-TLB Miss} – The translation is missing from both the \textit{micro-TLB} and \textit{macro-TLB}, necessitating a complete \ac{PTW}, incurring higher latencies.
\end{enumerate}

As demonstrated in both Figure~\ref{fig:smmu_contention_theo} and Figure~\ref{fig:smmu_contention_theo_diff_cache}, the performance penalty increases progressively from scenario (iii) to scenario (i). The most significant delay occurs when translations are missing in both TLB levels, requiring a full \ac{PTW} to proceed with the DMA transaction.

\mypara{Methodology.}
To assess the performance impact of shared \ac{SMMU} structures, we implemented a synthetic setup that emulates real-world scenarios using the Bao hypervisor. In this setup, an FPGA-based device performs DMA transactions as the benchmark, where we measure execution times. Meanwhile, a VM generates contention by issuing memory transactions through separate \ac{FPD}-\ac{DMA} channels. Specifically, the FPGA-based device accesses DDR memory via a \ac{DMA}-based IP connected through the S\_AXI\_HP3\_FPD interface, while the VM concurrently issues read operations from DDR memory using the \ac{FPD}-\ac{DMA}. As shown in Figure~\ref{fig:iommu_contention_setup}, both devices share the same \ac{TBU}, specifically $TBU\ 5$.

While adding more channels to the FPGA's \ac{DMA} controller could simulate contention, this approach risks introducing interference at the controller level, as multiple channels would compete for the same internal resources. Instead, by leveraging a separate \ac{FPD}-\ac{DMA} channel controlled by the VM, we create controlled contention without affecting the FPGA DMA controller itself. This allows us to isolate the impact on the \ac{SMMU} structures, ensuring a clearer understanding of performance degradation at both the \ac{TBU} and \ac{TCU} levels.

To further evaluate the impact of varying operational conditions, we adjust the FPGA clock frequency in our experiments. The FPGA DMA operates at 100MHz in the base case, but this frequency is lower compared to other DMA devices integrated into the SoC (e.g., the FPD-DMA, which operates at 600MHz). By varying the FPGA clock frequency, we aim to simulate more realistic scenarios.
Specifically, we increased the frequency to 150MHz and 300MHz to model the behavior of high-throughput DMA engines (e.g., PCIe or FPD-DMA) and observe how performance scales with increasing translation request rates. This change brings our controlled setup closer to real-world scenarios, while still isolating the effects within the IOMMU’s internal structures.
%This approach enables us to extrapolate how the performance of the FPGA DMA device might behave under typical operating conditions, where more frequent address translations occur due to higher clock speeds. We specifically vary the frequency to 150MHz and 300MHz, helping us understand the scaling of performance with increasing clock speeds and allowing for a comparison to non-CPU bus master devices in real-world use cases.

The evaluation consists of two configurations designed to explore the performance implications of shared \ac{SMMU} structures under the interference scenarios introduced earlier. These configurations directly assess contention at the micro-TLB (\ac{TBU}) and maco-TLB (\ac{TCU}) levels (e.g., by varying payload sizes), corresponding to the scenarios \textit{interf\_tbu} and \textit{interf\_tcu}, respectively.
In the first configuration (\textit{interf\_tbu}), we assess contention at the \ac{TBU}, as both the \ac{FPD}-\ac{DMA} and the S\_AXI\_HP3\_FPD interface are routed through \ac{TBU}5, as shown in Figure~\ref{fig:platform_overview}. This setup reflects real-world scenarios in which multiple devices rely on the same \ac{TBU} for address translations, leading to interference at the IOTLB. By analyzing this scenario, we aim to quantify the performance impact when \ac{DMA} devices contend for the same \ac{TBU} resources, especially when translations are not cached locally and require further interaction with the \ac{SMMU}.
The second configuration (\textit{interf\_tcu}) targets contention at the \ac{SMMU}'s \ac{TCU}, specifically its caching hierarchy. In this setup, concurrent access patterns from multiple devices stress the \ac{SMMU} caches, creating scenarios where frequent \ac{PTW} operations are required to perform address translations. Such conditions are common in use cases where various devices or VMs issue frequent and competing translation requests. This configuration allows us to evaluate how the \ac{TCU} handles high contention, focusing on the latency differences when translations are either cached in the \textit{macro-TLB} or entirely absent, generating \ac{PTW} request.

\definecolor{solo}{RGB}{94. 129. 172}
\definecolor{tbu}{RGB}{46. 52. 64}
\definecolor{tcu}{RGB}{191. 97. 106}

\arrayrulecolor{black!60} % Lighten the table borders
\begin{figure*}[t]

\centering
\includegraphics[width=.95\linewidth]{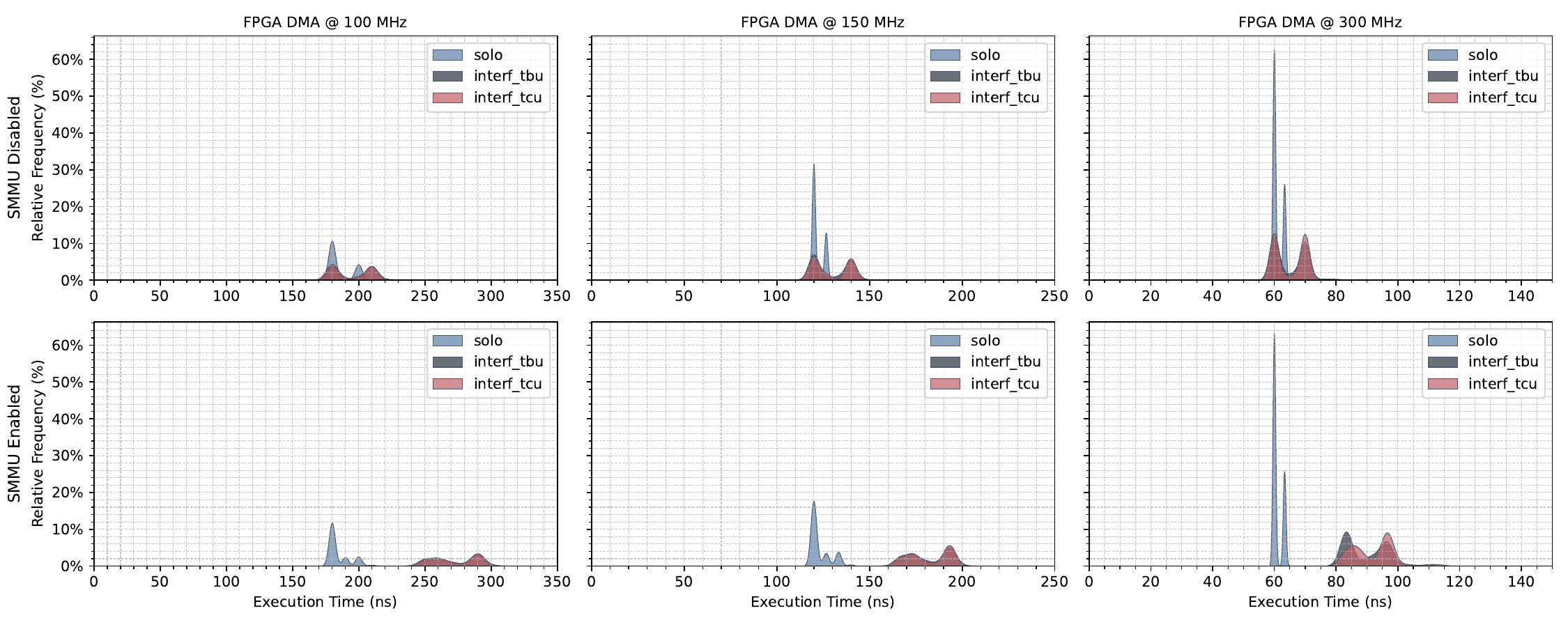}

\resizebox{.95\linewidth}{!}{%
\begin{tabular}{ccccclccclccclccc}
\hhline{-----------------}
\multirow{2}{*}{\begin{tabular}[c]{@{}c@{}}\textbf{SMMU }\\\textbf{Enabled }\end{tabular}} & \multirow{2}{*}{\begin{tabular}[c]{@{}c@{}}\textbf{FPGA }\\\textbf{Frequency }\end{tabular}} & \multicolumn{4}{c}{\begin{tabular}[c]{@{}c@{}}\textbf{Average Execution}\\\textbf{Time (ns) }\end{tabular}} & \multicolumn{4}{c}{\begin{tabular}[c]{@{}c@{}}\textbf{Peak Density }\\\textbf{Time (ns) }\end{tabular}} & \multicolumn{4}{c}{\begin{tabular}[c]{@{}c@{}}\textbf{Max Execution}\\\textbf{Time (ns) }\end{tabular}} & \multicolumn{3}{c}{\begin{tabular}[c]{@{}c@{}}\textbf{Min Execution }\\\textbf{Time (ns) }\end{tabular}} \\
                                                                                           &                                                                                              & {\cellcolor{solo!20}}\textbf{\textit{solo}} & {\cellcolor{tbu!20}}\textbf{\textit{interf\_tbu}} & {\cellcolor{tcu!20}}\textbf{\textit{interf\_tcu}} &  & {\cellcolor{solo!20}}\textbf{\textit{solo}} & {\cellcolor{tbu!20}}\textbf{\textit{interf\_tbu}} & {\cellcolor{tcu!20}}\textbf{\textit{interf\_tcu}} &  & {\cellcolor{solo!20}}\textbf{\textit{solo}} & {\cellcolor{tbu!20}}\textbf{\textit{interf\_tbu}} & {\cellcolor{tcu!20}}\textbf{\textit{interf\_tcu}} &  & {\cellcolor{solo!20}}\textbf{\textit{solo}} & {\cellcolor{tbu!20}}\textbf{\textit{interf\_tbu}} & {\cellcolor{tcu!20}}\textbf{\textit{interf\_tcu}}  \\ 
\hhline{-----------------}
\multirow{3}{*}{Disabled} 
    & 100MHz 
        & {\cellcolor{solo!20}}\begin{tabular}[c]{@{}c@{}}  185.71  \\\textbf{(baseline)}\end{tabular} 
        & {\cellcolor{tbu!20}}\begin{tabular}[c]{@{}c@{}}   194.79  \\\textbf{(1.05x)}\end{tabular} 
        & {\cellcolor{tcu!20}}\begin{tabular}[c]{@{}c@{}}   195.25  \\\textbf{(1.05x)}\end{tabular} &  
        & {\cellcolor{solo!20}}\begin{tabular}[c]{@{}c@{}}  180.00  \\\textbf{(baseline)}\end{tabular} 
        & {\cellcolor{tbu!20}}\begin{tabular}[c]{@{}c@{}}   180.10  \\\textbf{(1.00x)}\end{tabular} 
        & {\cellcolor{tcu!20}}\begin{tabular}[c]{@{}c@{}}   180.05  \\\textbf{(1.00x)}\end{tabular} &  
        & {\cellcolor{solo!20}}\begin{tabular}[c]{@{}c@{}}  200.00  \\\textbf{(baseline)}\end{tabular} 
        & {\cellcolor{tbu!20}}\begin{tabular}[c]{@{}c@{}}   230.00  \\\textbf{(1.15x)}\end{tabular} 
        & {\cellcolor{tcu!20}}\begin{tabular}[c]{@{}c@{}}   230.00  \\\textbf{(1.15x)}\end{tabular} &  
        & {\cellcolor{solo!20}}\begin{tabular}[c]{@{}c@{}}  180.00  \\\textbf{(baseline)}\end{tabular} 
        & {\cellcolor{tbu!20}}\begin{tabular}[c]{@{}c@{}}   180.00  \\\textbf{(1.00x)}\end{tabular} 
        & {\cellcolor{tcu!20}}\begin{tabular}[c]{@{}c@{}}   180.00  \\\textbf{(1.00x)}\end{tabular}  \\
    & 150MHz 
        & {\cellcolor{solo!20}}\begin{tabular}[c]{@{}c@{}}  121.93  \\\textbf{(baseline)}\end{tabular} 
        & {\cellcolor{tbu!20}}\begin{tabular}[c]{@{}c@{}}   129.10  \\\textbf{(1.06x)}\end{tabular} 
        & {\cellcolor{tcu!20}}\begin{tabular}[c]{@{}c@{}}   129.66  \\\textbf{(1.06x)}\end{tabular} &  
        & {\cellcolor{solo!20}}\begin{tabular}[c]{@{}c@{}}  120.00  \\\textbf{(baseline)}\end{tabular} 
        & {\cellcolor{tbu!20}}\begin{tabular}[c]{@{}c@{}}   120.05  \\\textbf{(1.00x)}\end{tabular} 
        & {\cellcolor{tcu!20}}\begin{tabular}[c]{@{}c@{}}   120.11  \\\textbf{(1.00x)}\end{tabular} &  
        & {\cellcolor{solo!20}}\begin{tabular}[c]{@{}c@{}}  126.67  \\\textbf{(baseline)}\end{tabular} 
        & {\cellcolor{tbu!20}}\begin{tabular}[c]{@{}c@{}}   146.67  \\\textbf{(1.16x)}\end{tabular} 
        & {\cellcolor{tcu!20}}\begin{tabular}[c]{@{}c@{}}   146.67  \\\textbf{(1.16x)}\end{tabular} &  
        & {\cellcolor{solo!20}}\begin{tabular}[c]{@{}c@{}}  120.00  \\\textbf{(baseline)}\end{tabular} 
        & {\cellcolor{tbu!20}}\begin{tabular}[c]{@{}c@{}}   120.00  \\\textbf{(1.00x)}\end{tabular} 
        & {\cellcolor{tcu!20}}\begin{tabular}[c]{@{}c@{}}   120.00  \\\textbf{(1.00x)}\end{tabular}  \\
    & 300MHz 
        & {\cellcolor{solo!20}}\begin{tabular}[c]{@{}c@{}}  60.98  \\\textbf{(baseline)}\end{tabular} 
        & {\cellcolor{tbu!20}}\begin{tabular}[c]{@{}c@{}}   64.98  \\\textbf{(1.07x)}\end{tabular} 
        & {\cellcolor{tcu!20}}\begin{tabular}[c]{@{}c@{}}   65.04  \\\textbf{(1.07x)}\end{tabular} &  
        & {\cellcolor{solo!20}}\begin{tabular}[c]{@{}c@{}}  60.00  \\\textbf{(baseline)}\end{tabular} 
        & {\cellcolor{tbu!20}}\begin{tabular}[c]{@{}c@{}}   60.04  \\\textbf{(1.00x)}\end{tabular} 
        & {\cellcolor{tcu!20}}\begin{tabular}[c]{@{}c@{}}   60.01  \\\textbf{(1.00x)}\end{tabular} &  
        & {\cellcolor{solo!20}}\begin{tabular}[c]{@{}c@{}}  63.33  \\\textbf{(baseline)}\end{tabular} 
        & {\cellcolor{tbu!20}}\begin{tabular}[c]{@{}c@{}}   80.00  \\\textbf{(1.26x)}\end{tabular} 
        & {\cellcolor{tcu!20}}\begin{tabular}[c]{@{}c@{}}   73.33  \\\textbf{(1.16x)}\end{tabular} &  
        & {\cellcolor{solo!20}}\begin{tabular}[c]{@{}c@{}}  60.00  \\\textbf{(baseline)}\end{tabular} 
        & {\cellcolor{tbu!20}}\begin{tabular}[c]{@{}c@{}}   60.00  \\\textbf{(1.00x)}\end{tabular} 
        & {\cellcolor{tcu!20}}\begin{tabular}[c]{@{}c@{}}   60.00  \\\textbf{(1.00x)}\end{tabular}  \\
\hhline{-----------------}
\multirow{3}{*}{Enabled} 
    & 100MHz 
        & {\cellcolor{solo!20}}\begin{tabular}[c]{@{}c@{}}  184.82  \\\textbf{(baseline)}\end{tabular} 
        & {\cellcolor{tbu!20}}\begin{tabular}[c]{@{}c@{}}   272.29  \\\textbf{(1.47x)}\end{tabular} 
        & {\cellcolor{tcu!20}}\begin{tabular}[c]{@{}c@{}}   273.35  \\\textbf{(1.48x)}\end{tabular} &  
        & {\cellcolor{solo!20}}\begin{tabular}[c]{@{}c@{}}  180.00  \\\textbf{(baseline)}\end{tabular} 
        & {\cellcolor{tbu!20}}\begin{tabular}[c]{@{}c@{}}   289.89  \\\textbf{(1.61x)}\end{tabular} 
        & {\cellcolor{tcu!20}}\begin{tabular}[c]{@{}c@{}}   289.94  \\\textbf{(1.61x)}\end{tabular} &  
        & {\cellcolor{solo!20}}\begin{tabular}[c]{@{}c@{}}  210.00  \\\textbf{(baseline)}\end{tabular} 
        & {\cellcolor{tbu!20}}\begin{tabular}[c]{@{}c@{}}   300.00  \\\textbf{(1.43x)}\end{tabular} 
        & {\cellcolor{tcu!20}}\begin{tabular}[c]{@{}c@{}}   300.00  \\\textbf{(1.43x)}\end{tabular} &  
        & {\cellcolor{solo!20}}\begin{tabular}[c]{@{}c@{}}  180.00  \\\textbf{(baseline)}\end{tabular} 
        & {\cellcolor{tbu!20}}\begin{tabular}[c]{@{}c@{}}   250.00  \\\textbf{(1.39x)}\end{tabular} 
        & {\cellcolor{tcu!20}}\begin{tabular}[c]{@{}c@{}}   250.00  \\\textbf{(1.39x)}\end{tabular}  \\
    & 150MHz 
        & {\cellcolor{solo!20}}\begin{tabular}[c]{@{}c@{}}  123.18  \\\textbf{(baseline)}\end{tabular} 
        & {\cellcolor{tbu!20}}\begin{tabular}[c]{@{}c@{}}   182.07  \\\textbf{(1.48x)}\end{tabular} 
        & {\cellcolor{tcu!20}}\begin{tabular}[c]{@{}c@{}}   182.44  \\\textbf{(1.48x)}\end{tabular} &  
        & {\cellcolor{solo!20}}\begin{tabular}[c]{@{}c@{}}  120.00  \\\textbf{(baseline)}\end{tabular} 
        & {\cellcolor{tbu!20}}\begin{tabular}[c]{@{}c@{}}   193.26  \\\textbf{(1.61x)}\end{tabular} 
        & {\cellcolor{tcu!20}}\begin{tabular}[c]{@{}c@{}}   193.28  \\\textbf{(1.61x)}\end{tabular} &  
        & {\cellcolor{solo!20}}\begin{tabular}[c]{@{}c@{}}  140.00  \\\textbf{(baseline)}\end{tabular} 
        & {\cellcolor{tbu!20}}\begin{tabular}[c]{@{}c@{}}   200.00  \\\textbf{(1.43x)}\end{tabular} 
        & {\cellcolor{tcu!20}}\begin{tabular}[c]{@{}c@{}}   193.33  \\\textbf{(1.38x)}\end{tabular} &  
        & {\cellcolor{solo!20}}\begin{tabular}[c]{@{}c@{}}  120.00  \\\textbf{(baseline)}\end{tabular} 
        & {\cellcolor{tbu!20}}\begin{tabular}[c]{@{}c@{}}   166.67  \\\textbf{(1.39x)}\end{tabular} 
        & {\cellcolor{tcu!20}}\begin{tabular}[c]{@{}c@{}}   166.67  \\\textbf{(1.39x)}\end{tabular}  \\
    & 300MHz 
        & {\cellcolor{solo!20}}\begin{tabular}[c]{@{}c@{}}  60.96  \\\textbf{(baseline)}\end{tabular} 
        & {\cellcolor{tbu!20}}\begin{tabular}[c]{@{}c@{}}   89.77  \\\textbf{(1.47x)}\end{tabular} 
        & {\cellcolor{tcu!20}}\begin{tabular}[c]{@{}c@{}}   92.27  \\\textbf{(1.51x)}\end{tabular} &  
        & {\cellcolor{solo!20}}\begin{tabular}[c]{@{}c@{}}  60.00  \\\textbf{(baseline)}\end{tabular} 
        & {\cellcolor{tbu!20}}\begin{tabular}[c]{@{}c@{}}   83.36  \\\textbf{(1.39x)}\end{tabular} 
        & {\cellcolor{tcu!20}}\begin{tabular}[c]{@{}c@{}}   96.61  \\\textbf{(1.61x)}\end{tabular} &  
        & {\cellcolor{solo!20}}\begin{tabular}[c]{@{}c@{}}  63.33  \\\textbf{(baseline)}\end{tabular} 
        & {\cellcolor{tbu!20}}\begin{tabular}[c]{@{}c@{}}   110.00  \\\textbf{(1.74x)}\end{tabular} 
        & {\cellcolor{tcu!20}}\begin{tabular}[c]{@{}c@{}}   113.33  \\\textbf{(1.79x)}\end{tabular} &  
        & {\cellcolor{solo!20}}\begin{tabular}[c]{@{}c@{}}  60.00  \\\textbf{(baseline)}\end{tabular} 
        & {\cellcolor{tbu!20}}\begin{tabular}[c]{@{}c@{}}   83.33  \\\textbf{(1.39x)}\end{tabular} 
        & {\cellcolor{tcu!20}}\begin{tabular}[c]{@{}c@{}}   83.33  \\\textbf{(1.39x)}\end{tabular}  \\
\hhline{-----------------}
\end{tabular}
}
\caption{Performance impact of SMMU contention.}
\label{fig:smmu_performance}

\end{figure*}

\mypara{Benchmarking Workload.} We deployed an IP core on the FPGA designed to issue DMA transactions in a circular pattern, enabling precise control over workload parameters. The core allows the configuration of (i) the payload size for each \ac{DMA} transaction, (ii) the number of memory pages accessed in sequence before wrapping around, and (iii) the base address from which transactions begin. To ensure consistency across experiments, we used a fixed page size of 4KB.
Although we considered using superpages (e.g., 2 MB pages) for potential performance gains, this configuration was discarded due to the complexities they introduce, such as reduced TLB efficiency, higher miss rates, and less flexible memory allocation. Moreover, superpages precludes the use of cache coloring, which helps mitigate interference at the \ac{LLC}.
For our experiments, the IP core was configured to issue DMA transactions in a circular manner, accessing the first address of 63 consecutive memory pages before wrapping around to the initial position. Each entry in this circular pattern corresponds to a DMA transaction rather than an explicit read operation. This results in 63 entries allocated in the micro-\ac{TLB} for read operations and one entry for write operations. The payload size for each transaction was set to 16 bytes, ensuring a lightweight yet consistent workload.
This setup was designed to focus exclusively on measuring the peak performance of \ac{DMA} transfers while minimizing interference from other shared hardware resources (e.g., DDR memory).
%By isolating these factors, we obtained a clear assessment of the \ac{SMMU}'s performance limits under controlled and well-defined conditions.

\mypara{Interference Workload.} To evaluate the impact of contention on the \ac{SMMU}, we deployed a baremetal VM running on top of the Bao hypervisor, configured to use eight \ac{DMA} channels for exerting stress on both the \ac{TBU} and \ac{TCU} caches. To maximize the volume of transaction requests, all four available CPUs on the platform are allocated to the VM. Each CPU manages two distinct \ac{DMA} channels, with each channel accessing $N/4$ unique pages, where $N$ is the total number of pages targeted by the application.
A key characteristic of this workload is that each of the eight \ac{DMA} channels operates with a unique Stream-ID. This differs from the FPGA IP workload, where only a single translation entry is needed for the page being accessed. In contrast, the interference workload's use of eight distinct Stream-IDs requires eight separate translation entries, even when all \ac{DMA} channels write to the same memory page. This design amplifies contention at the \ac{TBU}, as the independent translations for each Stream-ID add additional pressure on its caching resources.
To analyze the impact of this contention, we configured two scenarios. First, with $N=56$, the workload targets the \ac{TBU} (\textit{interf\_tbu}), using 64 total entries, a configuration to stress the micro-\ac{TLB}. Second, with $N=1992$, the workload shifts focus to the \ac{TCU} cache (\textit{interf\_tcu}), enabling the assessment of the \ac{SMMU}’s behavior under significantly higher contention levels.
These values were selected based on the IOMMU’s architectural details (as discussed in Section \ref{sec:cache_depth_measurements}), to induce contention directly within the IOMMU structure. To isolate IOMMU contention effects, a fixed and small payload size was used across all configurations. This approach minimizes interference from main memory and system bus activity, keeping performance degradation primarily due to translation-related contention.

\subsection{Latency of DMA Transactions with SMMU Disabled}

\mypara{Solo Execution.}
When the SMMU is disabled, the solo execution case serves as a baseline measurement for pure DMA performance without address translation overhead. The FPGA-based DMA engine directly accesses physical memory without intervention from the SMMU, ensuring that latency is only influenced by factors such as system interconnect and memory access times.
At a 100MHz FPGA frequency, the average execution time per DMA transaction is 185.71ns. This time scales linearly with the FPGA clock frequency, meaning that at 150MHz, the execution time reduces to 123.18ns, and at 300MHz, it reaches 60.96ns. The near-perfect scaling confirms that in a solo environment, execution time is tightly coupled with the clock frequency, with minimal external sources of variation.
The peak relative frequency (i.e., the most commonly observed execution time) aligns closely with the average execution time across all tested frequencies, reinforcing the predictability of solo execution. Furthermore, the minimum execution time remains unchanged, indicating a highly deterministic operation.

\mypara{Interference Scenarios.}
Under interference conditions (\textit{interf\_tbu} and \textit{interf\_tcu}), execution times increase slightly. At 100MHz, the average execution time rises to 194.79ns for \textit{interf\_tbu} and 195.25ns for \textit{interf\_tcu}, reflecting a 1.05x overhead compared to the solo case. A similar pattern is observed at higher FPGA frequencies, where interference induces only minor variations in the average execution time.
However, the maximum execution time fluctuates, with the highest overhead occurring in the \textbf{\textit{interf\_tbu}} scenario at 300MHz, reaching 126.67ns (1.26x higher when compared to the baseline). This suggests occasional contention at the memory subsystem and system bus, as further analyzed in Figure \ref{fig:smmu_performance}. Importantly, these effects primarily impact the latency of individual transactions rather than the overall throughput, indicating that while interference causes higher delays in specific transactions, it does not significantly affect the overall bandwidth.

%\begin{figure*}[t]
%    \centering
%    \includegraphics[width=.95\linewidth]{00_Figures/smmu_interference_w_freqs.pdf}
%    \caption{IOMMU Contention}
%    \label{fig:iommu_conention}
%\end{figure*}

\subsection{Latency of DMA Transactions with SMMU Enabled}

\mypara{Solo Execution.}
When the SMMU is enabled, the solo execution scenario assesses the impact of address translation overhead in an environment free from contention. Unlike the SMMU-disabled case, each DMA transaction must now pass through the IOMMU translation process, introducing potential latency due to TLB lookups, PTW, and cache accesses.
Despite these additional steps, solo execution latency remains nearly identical to the SMMU-disabled case. At 100MHz, the average execution time is 184.82ns, almost matching the 185.71ns observed without the SMMU. The execution time scales linearly with frequency, reducing to 123.18ns at 150MHz and 60.96ns at 300MHz, showing that in the absence of contention, address translations are efficiently handled by the SMMU's cache hierarchy, avoiding costly PTW operations.
Like in the SMMU-disabled case, the peak relative frequency remains stable across frequencies, confirming a highly predictable execution time in solo mode. Additionally, the minimum execution time is identical to the previous scenario, suggesting that best-case transactions are not affected by SMMU involvement.

\mypara{Interference Scenarios.} With the SMMU enabled, the impact of interference becomes significantly more pronounced. At 100MHz, the average execution time increases to 272.29ns for \textit{interf\_tbu} and 273.85ns for interf tcu, reflecting an overhead of 1.47x compared to \textit{solo} execution. Similarly, at 300MHz, interference results in a 1.51x overhead, highlighting the substantial performance cost introduced by contention at the SMMU. Unlike the SMMU-disabled case, minimum execution times are also affected, increasing by at least 1.39x across all frequencies. This suggests that even memory and system bus transactions, which are typically free from interference, experience contention under SMMU load, demonstrating that the SMMU consistently introduces a bottleneck in concurrent scenarios.
The maximum execution time reveals even more severe degradation. At 100MHz, the maximum execution time increases from 210ns (\textit{solo}) to 300ns in both interference scenarios, resulting in a 1.43x overhead. At 300MHz, the worst-case latency overhead reaches 1.79x in the \textit{interf\_tcu} scenario. This increase is attributed to contention at the SMMU’s TCU and TBU caches, which exacerbates transaction delays under high-load conditions. Figure \ref{fig:smmu_performance} further illustrates these findings, showing that SMMU contention primarily impacts worst-case latency, while average execution times remain more predictable. This highlights a key challenge: when the SMMU is heavily utilized, maintaining low-latency, deterministic transactions becomes increasingly difficult, posing significant challenges for real-time systems.

\subsection{Impact of SMMU contention on DMA Throughput}

\mypara{Methodology.} To evaluate the impact of SMMU contention on DMA throughput, we configured the FPGA IP to transfer varying payload sizes, ranging from the system bus width (128 bits, or 16B) to a memory page size (4KB). The DMA transfers were performed under both \textit{solo} execution and interference scenarios (\textit{interf\_tbu} and \textit{interf\_tcu}).

\begin{figure}[t]
    \centering
    \includegraphics[width=1\linewidth]{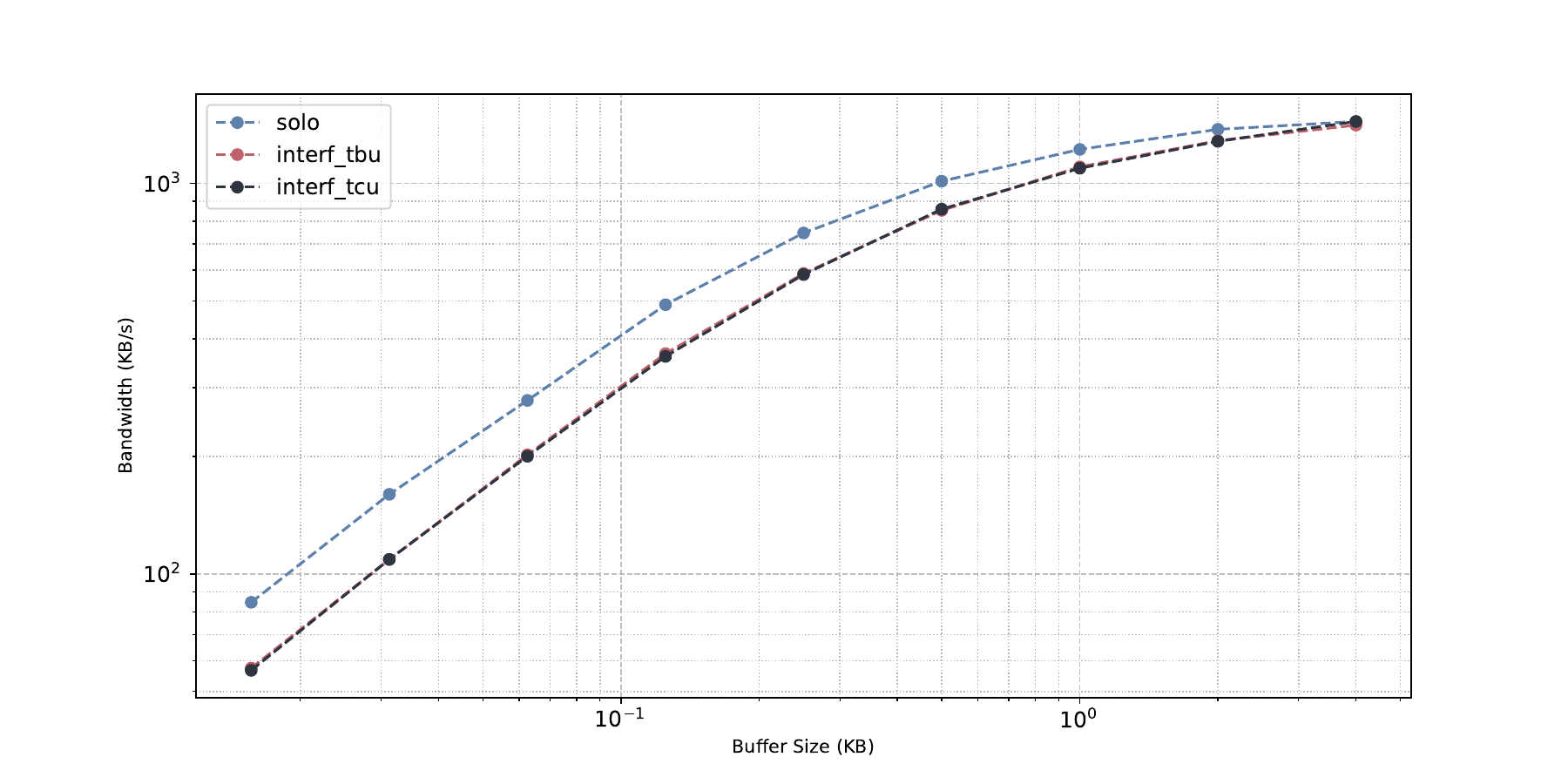}
    \caption{DMA transactions throughput}
    \label{fig:throughput_benchmark}
\end{figure}

\mypara{Solo Execution.} In the solo configuration, the DMA throughput scales predictably with the payload size, as shown in Figure \ref{fig:throughput_benchmark}. For small payloads (e.g., 16B), the bandwidth reaches approximately 84.6KB/s. As the payload size increases, the bandwidth improves significantly, reaching 1440KB/s for a 4KB payload. Empirical results show the linear relationship between payload size and bandwidth for smaller transfers.

\mypara{Interference Scenarios.} Under interference conditions (\textit{interf\_tbu} and \textit{interf\_tcu}), the DMA throughput is notably affected, particularly for smaller payloads: for a payload of 16B, the bandwidth decreases to from 84.6KB/s to 57.4KB/s and 56.7KB/s, respectively, corresponding to a reduction of approximately 32\% compared to the solo case. As the payload size increases, the impact of interference diminishes. For instance, at 1KB, the bandwidth for \textit{interf\_tbu} and \textit{interf\_tcu} reaches approximately 1102KB/s and 1094KB/s, respectively, which is only about 10\% lower than the solo configuration. At the maximum payload size of 4KB, the throughput for all configurations converges, with interference scenarios achieving bandwidths of up to 1410KB/s, closely matching the baseline.

\mypara{Key Observations.} The reduction in throughput due to interference is primarily caused by the overhead introduced by the SMMU during each DMA transaction. Empirical results show that this overhead remains constant at approximately 100ns, regardless of the payload size. Consequently, for smaller payloads, the overhead represents a significant portion of the total transaction time, leading to a substantial reduction in throughput. However, as the payload size increases, the total transaction time becomes dominated by the transfer of the payload itself (on the order of milliseconds), rendering the 100ns overhead relatively insignificant.

\section{Discussion and Future Directions}

This section examines the impact of IOMMU contention, emphasizing its effects on transaction latency and overall system performance. Additionally, we discuss key architectural factors influencing contention and outline future research to further quantify this issue across different platforms.

\subsection{IOMMU as a Source of Contention}
Our findings confirm that the IOMMU can act as a significant source of contention, introducing delays in memory transactions, particularly for smaller payload sizes. Experimental results indicate that lower-sized transactions can experience delays up to 1.79x due to contention in the IOMMU. This occurs because the overhead introduced by address translation and access control mechanisms within the IOMMU disproportionately affects transactions with shorter execution times. As a result, the IOMMU plays a non-negligible role in performance degradation for latency-sensitive applications, particularly those relying on frequent small transfers.
These latency increases are especially relevant in the context of real-time systems, as they may influence the ability of such systems to meet timing constraints. The observed delays could affect worst-case execution time (WCET) estimates, potentially requiring adjustments in system models to account for DMA contention. While a detailed exploration of real-time implications is beyond the scope of this work, the measured contention-induced latency increases can serve as a foundation for future real-time analysis.

\subsection{IOMMU Contention in Real-World Systems}
Analyzing contention in a real-world scenario, our experiments demonstrate that interference from TBU and TCU exhibit similar performance impacts on the evaluated platform. This equivalence is primarily attributed to the highly optimized caching structures present in the Arm SMMU, such as prefetch buffers and multi-level TLB hierarchies. The architectural refinements of SMMUv3, including the presence of Micro-TLBs, Macro-TLBs, and PTW caches, effectively mitigate excessive translation overhead by caching frequent translations and prefetching pages proactively. Consequently, contention effects are significantly absorbed, leading to comparable interference levels between the TBU and TCU. Furthermore, our findings suggest that the impact of IOMMU contention diminishes when considering larger payload transfers. The primary reason is that memory transaction latencies overshadow IOMMU processing times by orders of magnitude, causing the introduced overhead to be diluted over the total execution time. Thus, while smaller transfers suffer from substantial performance penalties, higher payload transactions experience relatively negligible impact from IOMMU contention.

\subsection{Mitigation Strategies for IOMMU Contention}
While the primary focus of this work was to characterize and quantify the impact of IOMMU contention, particularly in relation to DMA transaction latency, we recognize the importance of exploring mitigation strategies to alleviate such performance degradation. Although a full experimental evaluation of these techniques is outside the scope of this study, several promising approaches can be considered to reduce the impact of IOMMU-induced latency spikes.
One promising direction is the use of QoS regulators to manage bandwidth allocation across DMA stream-IDs. Recent IOMMU designs—such as those in RISC-V and Arm SMMUv3—integrate support for stream-ID-aware Quality-of-Service controls. These regulators can throttle or prioritize memory traffic on a per-stream basis, reducing interference and helping enforce bandwidth or latency constraints for critical tasks.
Another complementary approach involves static memory mapping and software-assisted page coloring. By mapping critical DMA workloads to predetermined page sets and avoiding page reuse across contending devices, designers can minimize pressure on IOMMU caching structures. Page coloring can also be used to partition IOMMU's caches to reduce interference between non-CPU bus masters, partitioning translation entries and improve locality.

\subsection{Future Directions}
The observations in this study highlight the need to further investigate IOMMU contention across different architectures. Since IOTLB distribution significantly affects contention, a broader experimental study on diverse architectures would be valuable.
In systems where multiple devices compete for shared address translation resources, the extent of contention may vary depending on the architectural design choices, such as the number and placement of IOTLBs.
Future research should also explore methods to mitigate IOMMU contention, such as dynamic translation prefetching strategies (e.g., preemptively fetching translation entries based on memory access patterns), adaptive TLB partitioning (e.g., allocating TLB resources differently based on workload characteristics), and hardware enhancements like hardware-managed TLB eviction (e.g., using specialized hardware to optimize TLB entry replacements). Additionally, evaluating contention effects in multi-tenant cloud environments or heterogeneous computing platforms can provide insights into how IOMMU behavior scales with an increasing number of concurrent memory-intensive workloads. By extending this study to a wider range of use cases, we can better understand and optimize IOMMU designs to improve system performance and efficiency.
\section{Related Work}

This section reviews prior work on IOMMU-related side channels and performance, focusing on security risks and architectural challenges relevant to contention and scalability.

\mypara{Side-channel attacks on IOMMU.} 
While IOMMUs aim to enforce isolation and protection for DMA transactions, recent studies show they also introduce new opportunities for side-channel attacks. Several works demonstrate how adversaries can exploit the IOTLB and peripheral interactions to bypass existing defenses.
Kim et al. introduced DevIOus \cite{kim2023devious}, a side-channel attack that leverages the IOTLB to extract sensitive information using DMA-capable PCIe devices such as GPUs and RDMA-enabled NICs. Unlike CPU-centric attacks, DevIOus operates entirely within the IOMMU domain, avoiding interference with CPU caches or TLBs.
Similarly, Markettos et al. presented Thunderclap \cite{markettos2019thunderclap}, which investigates vulnerabilities in OS-level IOMMU protections against DMA-based attacks. Despite widespread IOMMU deployment, they show how malicious peripherals can exploit shared memory regions and kernel interactions to hijack execution and extract data. Their findings are validated using a custom FPGA-based platform.
Expanding on these studies, Tiemann et al. \cite{Tiemann2023} introduced IOTLB-SC, identifying the IOTLB as a key leakage source in cloud systems. Using an FPGA accelerator, they uncover covert channels enabling unauthorized communication between peripherals, potentially leaking GPU-accelerated workloads. Their work highlights the growing relevance of IOTLB-based channels with emerging interconnects such as CXL and PCIe 5.0.
Lastly, Kim et al. \cite{kim2022pcie} analyzed the IOMMU’s role in PCIe-based side channels, focusing on RDMA-enabled network cards. By examining IOMMU translation behavior, they demonstrate how PCIe interfaces can serve as practical side-channel vectors in diverse systems.

\mypara{IOMMU performance.} %Understanding the performance implications of the IOMMU is critical for optimizing system efficiency, especially in heterogeneous environments that rely on DMA. 
Prior work has explored both architectural enhancements and empirical evaluations of IOMMU behavior \cite{ben2007price}.
Hur et al. \cite{hur2024performance} investigated the use of hashed page tables in the ARM SMMU, showing that this approach simplifies translation lookups and improves scalability in heterogeneous SoCs. By reducing lookup complexity, it can enhance performance for workloads with frequent DMA activity.
Paraskevas et al. \cite{paraskevas2020analysis} performed an experimental analysis of the ARM SMMU on a COTS platform, evaluating its behavior under real accelerator workloads. Their work provides useful insights into SMMU overheads but does not explore microarchitectural aspects such as micro- and macro-TLBs, or contention under co-allocated workloads.
Complementary to IOMMU-induced contention, Zini et al. \cite{zini2022profiling} studied memory contention caused by I/O devices on COTS platforms. They propose regulating bus traffic using COTS mechanisms like the Arm QoS-400 to improve performance predictability in \ac{MCSs}.
These studies highlight both the security risks and performance challenges introduced by the IOMMU, reinforcing the need for further investigation into contention, caching, and scalability in modern IOMMU designs.
\section{Conclusion}

The IOMMU is a critical component for memory protection and translation in modern computing systems, yet its impact on performance, particularly in mixed-criticality systems, has been largely overlooked. In this work, we analyzed contention effects within IOMMU structures, demonstrating that its shared nature can introduce interference that degrades the performance of time-sensitive workloads. Our findings show that this contention primarily affects small memory transactions, where translation delays are not diluted over long execution times. Additionally, based on the shared design principles of TBUs and TCUs across different architectures, we hypothesize that interference from these components would follow similar patterns, as their caching structures (e.g., prefetching and IOTLB designs) are based on common design principles, although further research is needed to confirm this.

These results highlight that the IOMMU is not only a security-relevant component but also a performance-sensitive element that must be carefully managed in heterogeneous MCSs. As system complexity increases, understanding and mitigating IOMMU-induced contention becomes essential to ensuring predictable performance. Future work should extend this analysis across different architectures, considering factors such as IOTLB distribution, resource sharing, and workload co-location. Addressing these challenges will be key to moving beyond the Bermuda Triangle of contention, toward a deeper and more comprehensive understanding of IOMMU behavior in complex systems.

\section*{Acknowledgment}

This work has been supported by FCT - Fundação para a Ciência e Tecnologia grants 2022.13378.BD, SFRH/BD/138660/2018.

\bibliographystyle{IEEEtran}
\bibliography{bibliography.bib}

\end{document}